\documentclass[pre,reprint,superscriptaddress,showpacs,amsmath,amssymb]{revtex4-1}

\usepackage[pdftex]{graphicx}
\usepackage{dcolumn}% Align table columns on decimal point
\usepackage{bm}% bold math
\usepackage{color}
\newcommand{\pdiff}[2]{\frac{\partial #1}{\partial #2}}
\newcommand{\odiff}[2]{\frac{d #1}{d #2}}
\newcommand{\cosT}{\cos \theta}

\begin{document}

\preprint{}

\title{
Drift instability in the motion of a fluid droplet with a
chemically reactive surface driven by Marangoni flow
}

\author{Natsuhiko Yoshinaga }
%\NY{
\email[E-mail: ]{yoshinaga@wpi-aimr.tohoku.ac.jp}
\affiliation{WPI-AIMR, Tohoku University, 
Sendai 980-8577, Japan}
%}

\author{Ken H. Nagai }
\affiliation{Department of Physics, Graduate School of Science,
The University of Tokyo, Tokyo 133-0033, Japan}

\author{Yutaka Sumino}
\affiliation{Faculty of Education, Aichi University of Education,
Aichi 448-8542, Japan}

\author{Hiroyuki Kitahata}
\affiliation{Department of Physics, Graduate School of Science, Chiba University, Chiba 263-8522, Japan}
\affiliation{PRESTO, JST, Saitama 332-0012, Japan}

\begin{abstract}
We theoretically derive the amplitude equations for a self-propelled
 droplet driven by Marangoni flow.
As advective flow driven by surface tension gradient is enhanced, the stationary state becomes unstable and
the droplet starts to move.
The velocity of the droplet is determined from a cubic nonlinear term
 in the amplitude equations.
The obtained critical point and the characteristic velocity are well supported by 
 numerical simulations. 
%The result is compared with numerical solution, and we found the good agreement.
\end{abstract}

%68.03.Cd

\maketitle

\section{Introduction}
Spontaneous motion or self-propulsion has been attracting attention in
recent decades because of its potential application to biological problems such
as cell motility \cite{BernheimGroswasser:2002,gucht:2005,Wiesner:2003,gerbal:2000,Cantat:2003}.
These intensive studies have stemmed from the fact that mechanical
properties of cells can be measured thanks to recent developments in visualization techniques \cite{lenz:2008a}.
In addition, several model experiments showing spontaneous
motion have been carried out \cite{dosSantos:1995,Toyota:2009,sumino:2005,nagai:2005,Thutupalli2011,thakur:2006}.
These systems consisted of relatively simple components such as oil
droplets in the water \cite{Toyota:2009}. 
Nevertheless, the droplets give the impression of being alive in that
they move spontaneously without being  pushed or
pulled, and they travel in straight lines, turn, and deform.

%The key questions are why a particle moves without external mechanical
%force (a force-free condition)
%and why it breaks symmetry and chooses one direction.

%The first question 
Motion in the absence of an external mechanical force 
%(force-free motion) 
has been discussed in terms of the Marangoni effect in which
a liquid droplet is driven by a surface tension gradient
\cite{Young:1959,levan:1981}.
The non-uniform surface tension can be controlled by an field variable
such as temperature and a chemical (typically surfactants) concentration
\cite{darhuber:2005}.
The mechanism is that the gradient induces convective flow inside and outside
of a droplet, which leads to motion of the droplet itself.
Similar flow and resulting motion are observed for a solid particle in
phoretic phenomena such as thermophoresis \cite{anderson:1989,jiang:2009}.
In both systems, objects are {\it swimming} in a fluid.
%This is similar to phoretic phenomena such as thermophoresis, which is
%the motion of particles toward a hotter or colder spot \cite{anderson:1989,jiang:2009},
%in the sense that flow around a particle or a droplet is generated
%autonomously so that it appears to be {\it swimming} in a fluid.

%There have also been several works on microswimmers in low-Reynolds
%hydrodynamics driven by irreversible cycle of conformational changes without
%external force \cite{najafi:2004,dreyfus:2005}.

The velocity of the above-mentioned motion is
%The essential quantities of interest for the force-free motion, such as the
%velocity of the droplet or particle, the velocity field surrounding it,
%and the force balance, 
reasonably well described using linear
theories \cite{Young:1959,anderson:1989,julicher:2009}.
This implies that the direction of motion is determined by some
asymmetry in the system
such as a temperature gradient (and/or a concentration gradient).
In the case of solid, an asymmetric particle has recently been created
by coating half of its surface with a different material.
Using this so-called Janus particle, the motion along a gradient created by the
particle itself, which is referred to as self-phoresis, was realized
\cite{Paxton:2004,jiang:2010,golestanian:2010}.
%Recently, the motion of a Janus particle along a gradient created by the
%particle itself, which is referred to as self-phoresis, was realized
%\cite{Paxton:2004,jiang:2010,golestanian:2010}.
%The Janus particle is asymmetric; half of its surface is coated with a
%different material.
The asymmetric field in this case is
not given externally but is created by consuming the energy supplied
uniformly from outside.
Nevertheless, linear theory still works sufficiently well since the particle has inherently asymmetric surface properties.

In contrast to the solid particle,
fluid droplets are dynamic and their surface properties cannot be fixed due to internal diffusion.
Motion in an isotropic system cannot be described
using a linear approach;
 it requires symmetry breaking arising from a
nonlinear term \cite{john:2008}. 
%Therefore the second question is more subtle in a isotropic system; a nonlinear term is necessary for motility.
%On the other hand,
In fact, spontaneous motion has been discussed 
using reaction-diffusion equations, which are
nonlinear partial differential equations,
%in the
%field of nonlinear dynamics in terms of 
 and is called as drift instability or drift
bifurcation
%, quite separately from soft
%matter physics 
 \cite{krischer:1994,schweitzer:1998, ohta:2009}.
%The instability is analyzed using reaction-diffusion equations, which are
%nonlinear partial differential equations \cite{ohta:2009}.
%In fact, Ohta {\it et al.} derived the mode equations of drift
%instability and deformation with symmetry
%consideration and from a particular reaction diffusion equation \cite{ohta:2009b,ohta:2009}.
%However, the coefficients of the equation and more importantly the
% velocity cannot be interpreted with physical constants such as a
% diffusion constant, viscosity, and so on.
Despite this, there has been few attempts to consider the mechanics and hydrodynamics of spontaneous motion.

In the present work, 
%we attempt to bridge the gap between two approaches:
%self-propulsion and drift bifurcation.
we derive amplitude equations showing drift bifurcation from a set of
equations for concentration fields taking hydrodynamics into
consideration.
All of the coefficients have
clear physical meanings, and can in principle be measured.
Our study is inspired by earlier pioneering works on the motion of 
reactive droplets \cite{ryazantsev:1985,rednikov:1989}.
While these studies mainly focused on linear stability and response to an
external force, 
our purpose is to derive equations containing nonlinear
terms
%in addition to inertia terms that can appear even under small systems.
and obtain the characteristic velocity of a droplet.

%Our system, particularly model III is similar to the pioneer work
%\cite{rednikov:1994} in which steady state of spontaneous motion was calculated.
%Our focus is not on the solution itself but derivation of {\it equation}
%satisfied by velocity of a drop.
%(advantage ... transparent )
%The advantage is that the coefficients are measurable in experiments.

\begin{figure}[hbt]
\begin{center}
\includegraphics[width=0.50\textwidth]{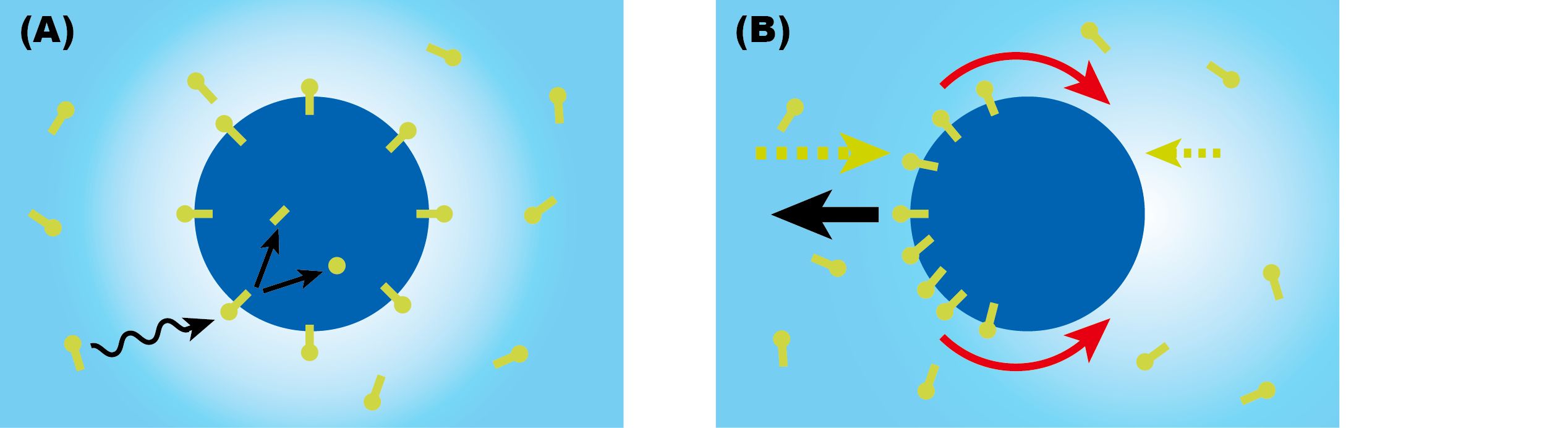}
\caption{\label{fig.schematic} 
(Color Online) Schematic illustration of the system in this study.
Surfactants dissolve in the outer fluid and some are adsorbed
 at the interface between the inner and the outer fluids. 
These surfactants reduce the surface tension of the droplet. 
(A) No droplet motion occurs for an isotropic distribution of surfactants.
(B) When the surfactant distribution becomes asymmetric, the flow (thin red arrows)
 occurs and the droplet starts to move in the direction of the thick black
 arrow.
The flux of surfactants are shown in broken arrows.
The background gradation represents surfactant concentration.
}
\end{center}
\end{figure}

\section{model}
We consider an axisymmetric system containing a spherical droplet in
a fluid which has an inner and/or outer surfactant concentration of
$c(r,\theta)$, and a velocity field of $ {\bf v} (r,\theta)= (v_{r}(r,\theta),v_{\theta}
(r, \theta))$ in the co-moving frame with the droplet \cite{sadhal:1996}.
Near the critical point of drift bifurcation, the velocity of the droplet is slow so that
${\bf v}(r,\theta)$ can be described by low-Reynolds hydrodynamics, that is, the Stokes equation
\begin{align}
 \eta \nabla^2 {\bf v} = \nabla p,
\end{align}
with the incompressible condition $\nabla \cdot {\bf v}=0$.
$\eta$ is the viscosity of the inner or outer fluid and $p$ is the pressure.
We assume a linear relationship between the concentration of
surfactants at the interface $\Gamma (\theta)$ and
surface tension 
\begin{align}
 \gamma(\theta) 
&= \gamma_0 + \gamma_c \Gamma(\theta)
\end{align}
using the surface tension $\gamma_0$ without surfactants.
The surfactant concentration at the interface can be expanded using Legendre polynomials as
\begin{align}
 \Gamma (\theta,t) &= 
\sum_{n=0}^{\infty} A_n(t) P_n (\cosT).
\end{align}
Here we restrict our attention to non-deformable droplets.
We consider only the $n=0$
and $n=1$ modes, and neglect the higher modes.
%Note that the higher modes ($n>2$) of the polynomials result in deformation.
The solution of the Stokes equation for Marangoni flow for a given
surface tension with an arbitrary distribution has been derived \cite{levan:1981,Kitahata:2011}.
It can be seen that the velocity of a droplet is proportional to the
first mode as
%The droplet velocity is
\begin{align}
 u = u_{1} A_1,
\label{setup_velocity_surfacetension}
\end{align}
where
\begin{align}
 u_{1} &=  
- \frac{2\gamma_c}{3(3\eta_i + 2 \eta_o)} ,
%\label{setup_velocity_viscosity}
\end{align}
and the subscripts ``i'' and ``o'' denote the inner
and outer fluid, respectively.
$\gamma_c$ is the strength of surface activity.
Since the surface tension is typically smaller for higher
concentrations of surfactants at the interface, $\gamma_c$ is negative and accordingly $u_{1} >
0$.
$u_1$ determines the strength of the chemomechanical coupling; 
%when $u_1=0$,
%an anisotropic distribution of surfactants does not create flow while
the flow field is sensitive to the anisotropy when $|u_1| $ is large.
Stronger coupling can be found for surfactants with higher surface
activity.
%The velocity fields $ {\rm v} (r,\theta)= (v_{r}(r,\theta),v_{\theta}
%(r, \theta))$ inside and outside a spherical drop are obtained under given
%surface tension of arbitrary distribution\cite{levan:1981,Kitahata:2011}.

The concentration of molecules adsorbed at the interface is in balance with
the bulk concentration field near the interface due to the
adsorption-desorption equilibrium as
\begin{align}
\alpha \Gamma (\theta)
=
 c (R,\theta),
\label{def_alpha}
\end{align}
where $\alpha$ is interpreted as the inverse of Henry's constant $K_{\rm
H}$ for adsorption equilibrium and has the dimensions of inverse length \cite{chang:1995}.
For a low surfactant concentration at the interface, $\alpha$ is simply described
as $k_d/k_a$ where $k_a$ and $k_d$ are the adsorption rate from bulk and desorption
rate from surface, respectively.
For this reason, $\alpha$ is not dimensionless but has the dimension of length.
%Therefore, 
For surfactants with higher surface activity $\alpha$ can be small, for
instance,
$\alpha \simeq
10^{-1}$ m$^{-1}$ \cite{chang:1995}
.
The concentration of surfactants at the interface can be expressed as\cite{sadhal:1996}
\begin{align}
 \pdiff{\Gamma}{t} 
+ v_{\theta} (R) \nabla_s \Gamma
= D_s \nabla_s^2 \Gamma
- \kappa_s \Gamma 
+ 
\left[
 D_o \pdiff{c}{r} 
- D_i  \pdiff{c}{r} 
\right]_{r=R}
,
\label{equation.gamma}
\end{align}
where the surface derivative is defined as $\nabla_s = (1/R)
\partial/\partial \theta$ for a sphere.
$D$ ($D_i$ and $D_o$) and $D_s$ are the bulk and surface diffusion constants, respectively.
%where the plus in the last term corresponds to outer bulk (the diffusion
%constant $D = D_o$ and the concentration 
%$c=c^o$) and the minus
%corresponds to inner bulk ($D = D_i$ and $c=c^i$).
The surfactants are reactive; molecules dissolved in the bulk are adsorbed
onto the interface, and after a characteristic time $\kappa_s^{-1}$
they lose their surfactant functionality, for instance by decomposing
into a head and a tail (see Fig.\ref{fig.schematic}).
We describe this by a linear reaction $-\kappa_s \Gamma$ with
a consumption rate $\kappa_s$.  
In this model, we implicitly assume addition and removal of surfactants
at the interface, which depend on the divergence of the two-dimensional
velocity fields.

We derive the amplitude equation near the onset of drift instability
where $A_1 \sim\epsilon$ is small so that $u \sim \epsilon$.
Our goal is to obtain the equation for the first mode 
\begin{align}
m \frac{d A_1}{dt} =
g A_1 + \epsilon \mathcal{F}_1 (A_1) 
+ \epsilon^2 \mathcal{F}_2 (A_1) + \cdots,
\end{align}
 with coefficients $m$ and $g$, and
some functions $\mathcal{F}_1$, $\mathcal{F}_2, \cdots$.
% which may or may not be dependent on $n=0$ and higher modes.
Taking (\ref{setup_velocity_surfacetension}) into consideration, this is equivalent to a
Landau-type equation for the droplet velocity
\begin{align}
\tilde{m} {d u}/{dt} =
\tilde{g} u + \epsilon \tilde{\mathcal{F}}_1 (u) 
+ \epsilon^2 \tilde{\mathcal{F}}_2 (u) + \cdots.
%\label{schematic.eq.velocity}
\end{align}
The basic idea is to eliminate the velocity and bulk concentration fields
in order to obtain a closed form of the equations for $A_1$.

We hereafter focus only on the surfactant concentration in the outer fluid and therefore drop the subscript
``o''. 
%We consider the three models in which the surfactants dissolve in (I) the outer
%bulk, (II) the inner bulk.
The two fluids under consideration could, for example, be water and oil, 
and the surfactants preferentially dissolve in either one or the other. 
%We also investigate the model III in which surfactants dissolve in the outer bulk without
%consumption.
%Here the concentration far from the interface is fixed using boundary condition.
%We will proceed the model I, but this approach is applied for the
%other models in a straightforward way.
The bulk concentration can be expressed using the Helmholtz equation
with advection, 
\begin{align}
 \pdiff{c}{t} + {\bf v} \cdot \nabla c
= D \nabla^2 c - \kappa (c - c_{\infty}).
\label{bulk_Helmholtz}
\end{align}
The model takes into account the supply of surfactants to the bulk in order to
maintain a constant concentration $c_{\infty}$ far from the interface.
The time scale is given by $\kappa$. 
We expand (\ref{bulk_Helmholtz}) around the critical point of drift
instability; the velocity of the droplet, $u$, or the P{\'{e}}clet number $R u/D$ is set as a small
parameter $\epsilon$.
We can solve this equation
perturbatively as
\begin{align}
 c(r,\theta) = c_{\infty} +
c^{(0)} (r,\theta) 
+ \epsilon c^{(1)} (r,\theta) 
%+ \epsilon^2 c^{(2)} (r,\theta) 
+ \cdots 
%\label{bulk_expansion}
\end{align}
with the boundary conditions at infinity
%\begin{align}
$c(\infty,\theta)=c_{\infty}$
%\end{align}
 and at the interface (see (\ref{def_alpha})). 
%At the lowest order,
For the orders of $\epsilon^0$ and $\epsilon$
%$\mathcal{O}(\epsilon^0)$ and $\mathcal{O}(\epsilon)$
,
(\ref{bulk_Helmholtz}) is expressed as
\begin{align}
 \odiff{c^{(0)}}{t}
&=
 D \nabla^2 c^{(0)} -
\kappa c^{(0)},
%\mbox{ for  \mathcal{O}(\epsilon^0)}
\label{bulk.zeroth}
\\
%\end{align}
% for $\mathcal{O}(\epsilon^0)$, 
%\begin{align}
 \odiff{c^{(1)}}{t}
+  {\bf v} \cdot \nabla c^{(0)}
&= D \nabla^2 c^{(1)} - \kappa c^{(1)}.
%\mbox{ for \mathcal{O}(\epsilon)}
\end{align}
% for $\mathcal{O}(\epsilon)$ and so on. 
The resulting $c(r)$ is then substituted back into
(\ref{equation.gamma}).
Due to the boundary condition (\ref{def_alpha}), the solution of $c(r,\theta)$ contains
the individual modes $A_n$ and coupled modes $A_n A_m$.
The nonlinear time evolution equations of $A_n$ are then obtained (see
(\ref{AmpEq0.scaling}) and (\ref{AmpEq1.scaling})).

\subsection{uniform distribution}
\begin{figure}[htb]
\begin{center}
\includegraphics[width=0.50\textwidth]{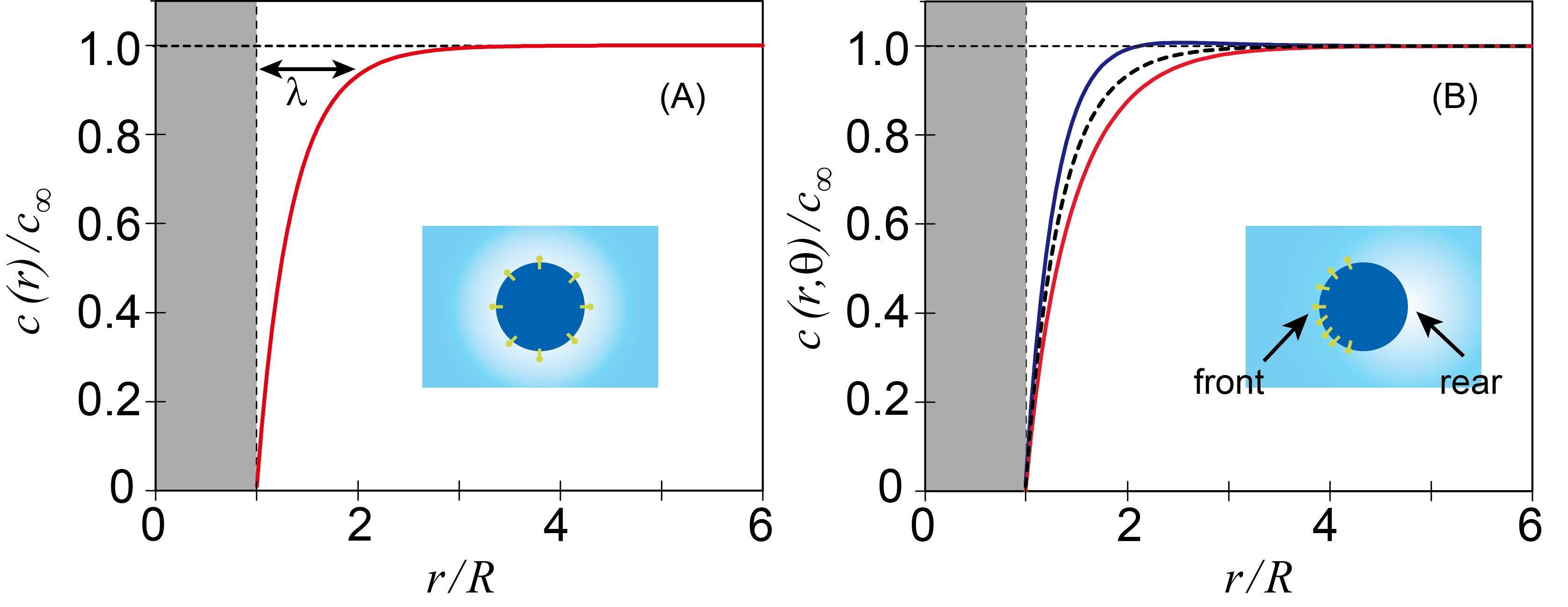}
\caption{
(Color Online) Distribution of bulk concentration field.
(A) Isotropic distribution when $\lambda=0.5$, $\alpha=0.01$, and
 $u_1/D=0$. (B) Anisotropic distribution with $u_1/D=2.0$. 
The blue
 (dark grey) line
 shows $c(r,\theta = 0)$ (front) and the red (light grey) line shows $c(r,\theta
 = \pi)$ (rear). 
The uniform
 distribution of (A) is shown in (B) as a dashed line.
\label{concentration_zero}
}
\end{center}
\end{figure}

We assume that the relaxation of the bulk concentration field is fast.
%It can be shown that the time derivative of $c(r)$ results in additional
%inertia-like terms in (\ref{AmpEq1.scaling}), although this is the
%beyond the scope of the present paper.
The zeroth order solution of (\ref{bulk.zeroth}) is then  
\begin{align}
  c^{(0)} (r,t) 
=
\left(
\alpha A_0 (t) - c_{\infty }
\right)
\frac{k_0 (r/\lambda)}{k_0 (R/\lambda)}
\label{solution_c_zero}
,
\end{align}
where $k_n(x)$ is an $n$th-order modified spherical Bessel function of the
second kind \cite{arfken:1968}.
%$ k_n(x) = \sqrt{2/(\pi x)} \mathcal{K}_{n+1/2}(x)$ using the
%$n$th-order modified Bessel function of second kind $\mathcal{K}_{n}(x)$
The result is plotted in Fig.\ref{concentration_zero}(A).
A steep gradient can be observed 
in the typical length scale
 $\lambda = \sqrt{D/\kappa}$.
%for $R \leq r \leq R+\lambda$, whereas the
%concentration saturates and becomes almost constant for $r> R+\lambda$.
The gradient is sustained by surface reaction characterized by
$\kappa_s$ in (\ref{equation.gamma}).
% and (\ref{equation.gamma.no.surfactant}).
For $R \gg \lambda$, the surface concentration is given by
\begin{align}
A_0 
& \simeq
\frac{c_{\infty}}{\kappa_s \lambda/D + \alpha},
\end{align}
leading to a gap $c_{\infty} - \alpha A_0$ between the concentration near the surface and at
infinity.
Since this gap is proportional to $\kappa_s$, the concentration gradient is driven by
surface reactions.

\subsection{Amplitude equations}
A weakly nonlinear analysis up to the order of $\epsilon^3$ shows 
%\cite{Note1}
%We next present the results of a weakly nonlinear analysis. 
%Details of the calculations will be reported elsewhere \cite{Note1}.
\begin{align}
 \odiff{A_0}{t} = &
- \kappa_s A_0 
+ \frac{\lambda + R}{\lambda R} D (c_{\infty} - \alpha A_0)
+  \Lambda^{(2)}_0 A_1^2
%+ \mathcal{O}(A_1^4)
,
\label{AmpEq0.scaling}
\\
%\end{align}
%\begin{align}
  \odiff{A_1}{t} = &
- \Lambda^{(1)}_1 \left(
1 - \frac{u_1}{u_1^*}
\right) A_1
 -
 \Lambda^{(3)}_1  A_1^3
%+ \mathcal{O}(A_1^4)
,
\label{AmpEq1.scaling}
\end{align}
where the coefficients are
\begin{align}
\Lambda^{(2)}_0 
&=
\Lambda_{02} u_{1} \alpha  
+ \Lambda_{03} \frac{ u_{1}^2}{D} 
(c_{\infty} - \alpha A_0)
 -   \frac{3u_{1}}{2R}, 
\label{Lambda0}
\\
%\end{align}
%\begin{align}
 \Lambda^{(1)}_1
&=
  \frac{2 D_s}{R^2} + \kappa_s 
+ \frac{D \alpha}{\lambda}, 
\label{Lambda1}
\\
%\end{align}
%\begin{align}
\Lambda^{(3)}_1 
&=
\Lambda_{13}
\frac{\alpha u_1^2 }{D} 
+ 
 \Lambda_{14}
\frac{u_{1} ^3}{D^2}
 (c_{\infty} - \alpha A_0),
\label{Lambda2}
\end{align}
with the coefficients $\Lambda_{ab}$ being dependent only on $\lambda$
and $R$.
The explicit forms of $\Lambda_{ab}$ are shown in the Appendix
(see (\ref{Lambda02})-(\ref{Lambda14})).
The critical point of the drift bifurcation occurs when the first
term on the right-hand side of  (\ref{AmpEq1.scaling}) changes its sign;
\begin{align}
u_1^* 
&=
\frac{  \frac{2 D_s}{R^2} + \kappa_s 
+ \frac{D \alpha}{\lambda}}{\Lambda_{12}   (c_{\infty} - \alpha A_0)}.
\end{align}
For $u_1 \leq u_1^{*}$, a stationary state is stable whereas it
becomes destabilized and the droplet moves for $u_1 > u_1^{*}$.
For $\alpha \ll \kappa_s \lambda /D$, the steady-state velocity of the droplet is given by
\begin{align}
 u \simeq 
%D \sqrt{\frac{\Lambda_{12}}{\Lambda_{14}}}
u_0
\sqrt{
1 - \frac{u_1^{*}}{u_1}
},
\label{theory.velcoity.alpha0}
\end{align}
where the characteristic velocity is $u_0 = \sqrt{D \kappa R
/\lambda}$ for $R \gg \lambda \gtrsim 0.01R$.

\begin{figure}[htb]
\begin{center}
\includegraphics[width=0.50\textwidth]{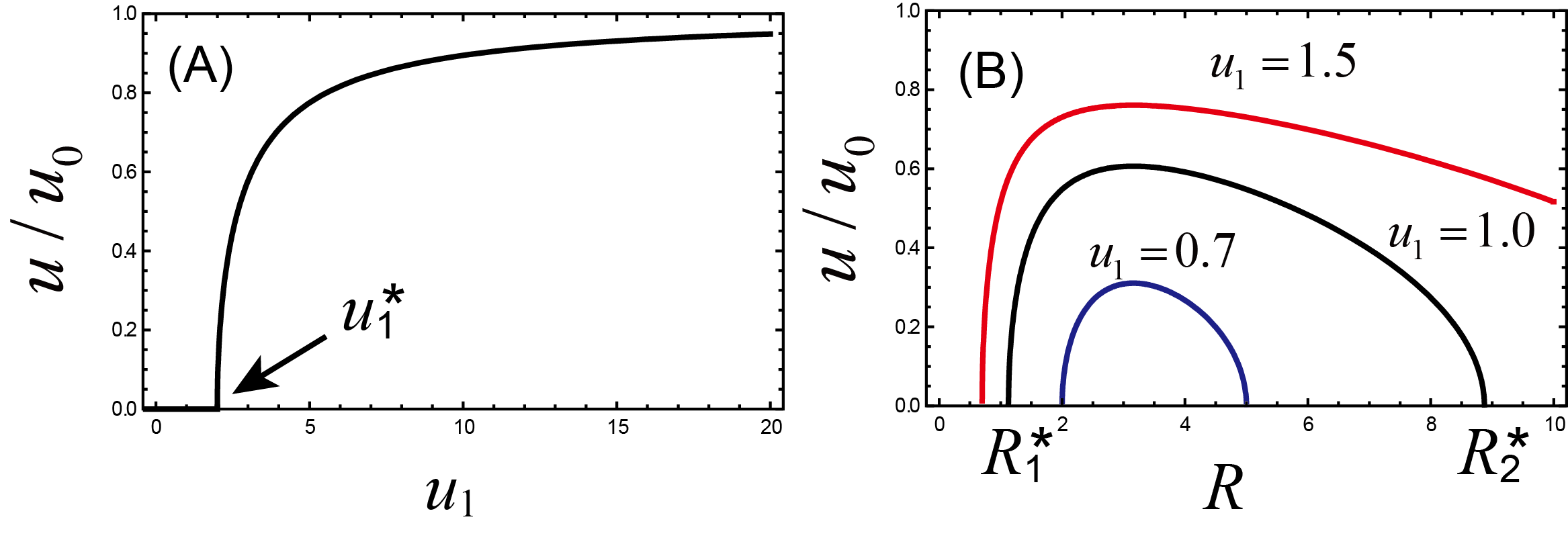}
\caption{\label{fig.bifurcation}          
(Color Online) Bifurcation diagram for spontaneous motion.
The bifurcation parameters are chosen to be $u_1$ (A) and $R$ (B). 
%(A) Model I and II show the supercritical bifurcation.
%(B) Model III shows subcritical bifurcation.
}
\end{center}
\end{figure}

%From an intuitive point of view, 
The instability can be explained as follows. 
First, small fluctuations in the surfactant concentration at the
interface give rise to a small
$A_1$, which induces convective flow around the droplet.
The flow then distorts the bulk concentration field through the
advection term.
Above the critical point, the distortion overcomes the relaxation
due to diffusion and amplifies the first mode $A_1$ leading
to further flow and motion of the droplet.
In fact, Fig. \ref{concentration_zero}(B) shows that the gradient in the bulk
concentration at the front of the droplet (relative to the direction of motion) is steeper than that at the rear.
This steeper gradient causes a larger flux from the bulk to the surface, and
thus leads to an inhomogeneous surface concentration.
Above the critical point, the
velocity increases with $u_1$ as in Fig.\ref{fig.bifurcation}.
In actual experiments, the size of a droplet may be the suitable parameter to vary. 
We find that there is an optimal droplet size for producing the highest
velocity 
%, and droplets that are significantly larger or smaller than
%this do not exhibit self-propulsion
(Fig.\ref{fig.bifurcation}B).
The two critical radii $R^{*}_1 \simeq D_s /c_{\infty} u_1 \lambda$ and
$R^{*}_2 \simeq c_{\infty} u_1 \lambda^2/(\lambda \kappa_s + D \alpha)$
arise from two stabilizing factors: surface diffusion and surface
reaction. 
Both of them are balanced with the effect of advection.
The size range for efficient self-propulsion increases with $u_1$.
The time evolution of the first mode below the critical point can be
expressed as
$%\begin{align}
 A_1 \sim e^{-t/\tau_{\rm relax}},
$%\end{align}
where the relaxation time is
\begin{align}
\tau_{\rm relax} 
&=
\left[
\frac{2D_s}{R^2} + \kappa_s + \frac{D \alpha}{\lambda}
\right]^{-1}
\left(1 - \frac{u_1}{u_1^*} \right)^{-1},
\label{theory.relaxationtime}
\end{align}
which diverges at $u_1 = u_1^*$.

In the linear term of (\ref{AmpEq1.scaling}), $\Lambda_1^{(1)}/u_1^{*} =
\Lambda_{12} (c_{\infty} - \alpha A_0)$, which corresponds to
(\ref{flux.first}) with (\ref{Lambda12}), destabilizes the stationary
state.
The physical origin of the destabilization is motion of the droplet.
This can be seen in the first bracket in the velocity in
radial direction (\ref{velocity.r}), which leads to the destabilization term.
The first term in the bracket $-u P_1(\cosT)$ corresponds to translational motion of the
droplet in the co-moving frame while the second term $u (R/r)^3 P_1(\cosT)$ arises from
convective flow around the droplet.
We investigated the contributions from both terms separately, and found
that two terms have opposite effects; the first term (translational
motion) destabilizes the stationary state while the latter (convection) stabilizes the
instability. 
The
instability is realized because the former always has stronger effect.

\section{Numerical simulations}

Numerical simulations are performed using spherical coordinates for
an axisymmetric three-dimensional system.
Both the radial and angular directions are discretized with $N +1$ mesh
points.
It is convenient to use the non-dimensionalized form of equations
(\ref{equation.gamma}) and (\ref{bulk_Helmholtz}).
\begin{align}
 \pdiff{\tilde{\Gamma} }{\tilde{t}} 
+ \frac{\tilde{v}_{\theta} (\tilde{R},\theta)}{\tau} \tilde{\nabla}_s \tilde{\Gamma}
&= l^2_s \tilde{\nabla}_s^2 \tilde{\Gamma}
-  \tilde{\Gamma} + \pdiff{\tilde{c}}{\tilde{r}},
\label{equation.gamma.norm}
\\
%\end{align}
%\begin{align}
\tau  \pdiff{\tilde{c}}{\tilde{t}} 
+ \tilde{{\bf v}} \cdot \tilde{\nabla} \tilde{c}
&= l^2 \tilde{\nabla}^2 \tilde{c} - \tilde{c},
\label{bulk_Helmholtz.norm}
\end{align}
where 
$
\tilde{\Gamma} 
= (R \kappa_s / D c_{\infty}) \Gamma,
$
$
\tilde{c} 
=
(c - c_{\infty}) / c_{\infty},
$
$\tilde{{\bf v}} = {\bf v} /\kappa R$,
$
\tilde{t} 
= \kappa_s t ,
$
$
\tilde{r} 
= r/R,
$
$
\tau 
= \kappa_s / \kappa,
$
$
l_s = \sqrt{D_s /\kappa_s R^2},
$ and 
$
l = \sqrt{D / \kappa R^2}
$.
The velocity field is also non-dimensionalized as
$
\tilde{u} 
= \tilde{u}_1 \tilde{A}_1,
$
where
$
 \tilde{u}_1
= (D c_{\infty} / \kappa \kappa_s R^2) u_1,
$
and
$
\tilde{A}_n 
=
(R \kappa_s / D c_{\infty}) A_n.
$
The boundary condition is rewritten as
$
\tilde{\alpha} \tilde{\Gamma} 
= \tilde{c}(1) + 1
$
with
$
\tilde{\alpha} 
= (D / \kappa_s R) \alpha
$.
%Then, the number of parameters is reduced as $l$,$l_s$, $\tau$,
%$\tilde{\alpha}$, and $\tilde{u}_1$.
We choose $\tilde{u}_1$ to be a bifurcation parameter, which induces
instability above a certain threshold.
$\tau$ is assigned a small value of 0.04.
We estimate the critical point from the relaxation time using (\ref{theory.relaxationtime}).

\begin{figure}[htb]
\begin{center}
\includegraphics[width=0.45\textwidth]{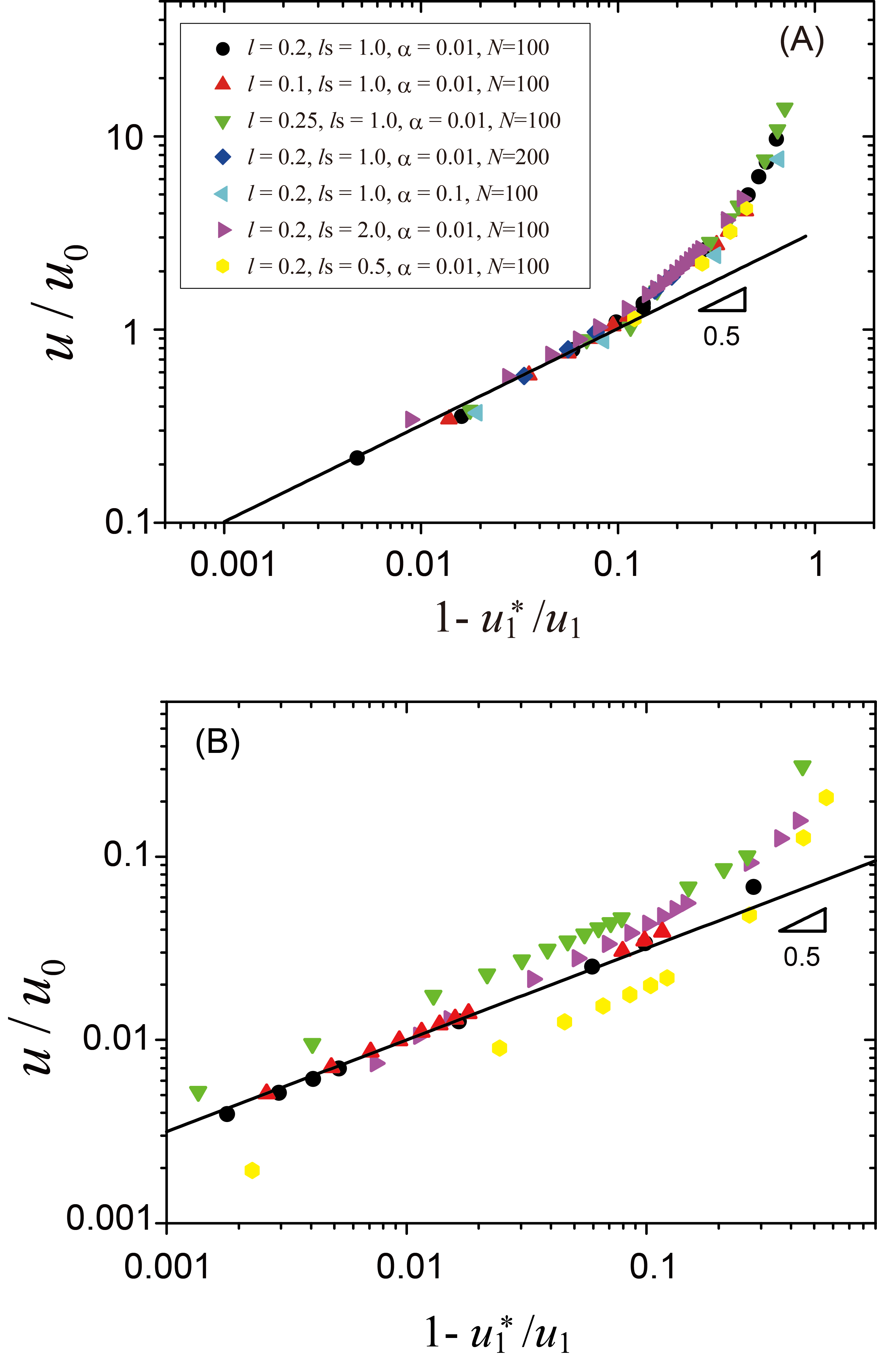}
\caption{\label{fig.numerics.bifurcation}          
(Color Online) Normalized velocity of a droplet without (A) and with
 (B) surface advection in (\ref{equation.gamma.norm}). The slope of the line is 0.5. 
}
\end{center}
\end{figure}

We estimate the critical point from the relaxation time above the
transition with (\ref{theory.relaxationtime}). 
Since the time evolution of $A_1$ decays exponentially, we estimate the
relaxation time by fitting the semi-log plot of $A_1$ as a function of time. 
From the $x$-intercept of the plot of
relaxation time as a function of $u_1$, we obtain the value of $u_1$ at the
critical point. 
The critical point weakly depends on the number of mesh points; for
instance, for $\tilde{l} =0.2$ and $\tilde{l}_s = 1.0$,
our theory predicts $\tilde{u}^{*}_1 = 1.03$ while the numerical
results show $\tilde{u}^{*}_1 = 1.09$ for $N=100$. 
As the mesh number is increased, the estimated critical point becomes
closer to the predicted value $\tilde{u}_1^{*}=1.08 $ for $N = 200$ and
$\tilde{u}_1^{*} = 1.04$ for $N = 400$. 
Nevertheless, Fig.~\ref{fig.numerics.bifurcation}
shows that the normalized plot using numerically estimated values does
not depend on the number of mesh points. 
We have mainly used $N = 100$ for saving computational time and
for earning data points.

The numerical results show the concentration distribution
around a droplet moving in the left direction \footnote{
See supplementary movie found in 
http://www.wpi-aimr.tohoku.ac.jp/~yoshinaga/index.html
}.
It can be seen that the concentration distribution around the droplet is asymmetric.
%The velocity is plotted as a function of $u_1$ in
%Fig.\ref{fig.numerics.bifurcation}. 
The droplet is stationary for small $u_1$ whereas it moves when
$u_1$ becomes larger.
Note that the direction of motion is determined by an initially introduced
small noise, and is therefore random.
%The relaxation time diverges at the critical point as in
%Fig. \ref{fig.numerics.relaxationtime}, and the numerical results are
%consistent with the theoretical result of (\ref{theory.relaxationtime})
%without any fitting parameters.
The velocity normalized by $u_0$ is plotted against the distance from
the critical point in
Fig. \ref{fig.numerics.bifurcation}.
Near the critical point the slope has a value of 0.5, which is
comparable to the analytical
result (\ref{theory.velcoity.alpha0}).
A bifurcation is observed both with and without the surface advection
term in (\ref{equation.gamma}).
The characteristic velocities
deviate slightly from the analytical results for some choices of
parameters when the surface advection is included.
This may be due to the effects of higher modes.
Without the surface advection, all of the data points lie on the same curve
irrespective of the parameters used.
%Surprisingly, the plots are collapsed even when the slope deviates from the
%theoretical line, suggesting that $u_0$ well characterizes the velocity.

\section{Summary and Remarks}
%\label{section:summary}
In summary, we derive amplitude equations for drift instability of a
droplet driven by Marangoni flow.
The critical point and the droplet velocity are calculated analytically, and
%We determine the characteristic velocity in terms of parameters
%with clear physical meaning.
good agreement is found with the results of numerical calculations.
%Although we focus on the systems with surfactants in this study, our
%model may also treat system without surfactants for example a reactive
%droplet in a fluid.
%%%%%%%%%%%%%%%%%%%%%%%%%
Our system is out of equilibrium due to the reaction at the interface by
which the supplied energy is consumed (see (\ref{equation.gamma})).
 This reaction maintains
a concentration gradient in the radial direction.
An additional key factor is the nonlinear advection term in the
bulk concentration field,
which leads to coupling between modes and breaks the symmetry of the
system.
The concentration gradient in the radial direction as
well as the
flux of surfactants onto the interface then  becomes
asymmetric. 
This leads to a surface tension gradient which results in motion. 
By contrast, surface advection is not
essential for motility.
%We focus on a system under low-Reynolds number where linear Stokes equation
%dominates. 
Despite the linear nature of the velocity fields associated with the
Stokes equation, we show that the addition of a
nonlinear term in the concentration field can lead to steady motion in an isotropic system.
Further studies are required in order to clarify the kinds of nonlinear
effects that are
necessary for motility.

%We expect this approach could be applied to other systems involving hydrodynamics.
%In fact, the numerical simulations done by Misbah {\it et al.} suggest the amplitude equation similar to
%that obtained in this paper \cite{kaoui:2009}.
Our model does not necessarily require the presence of surfactants.
For instance, a uniformly heated droplet or a droplet with a source of
chemicals can be tractable as the same manner with appropriate limits:
$\alpha \rightarrow 1$, $D_s \rightarrow 0$, and $\kappa_s \gg 1$.
%$u_1$ in this case is interpreted as the temperature dependence of the surface tension.
In this situation, (\ref{equation.gamma}) is equivalent to a
boundary condition for flux in the concentration field
$%\begin{align}
\left[
 D_o {\bf n} \cdot \nabla c 
- D_i {\bf n} \cdot \nabla c
\right]_{r=R}
- \kappa_s c(R) 
= 0,
$ 
%\label{equation.gamma.no.surfactant}
%\end{align}
where the first and second terms represent the flux from outside and inside of the droplet, respectively.
Here, the surface concentration $\Gamma$ independent of $c(r)$ does not
exist.
However, it is convenient to introduce a {\it virtual} surface
concentration 
%not only to achieve a unified framework
%with/without surfactants but also 
because velocity fields
are essentially created by the concentrations at the surface (see (\ref{setup_velocity_surfacetension})).
%This property is expected to appear most of transport phenomena driven
%by surface (or near surface) including capillary flow and electro- and
%thermo- phoresis.
It should also be stressed that similar results can be obtained using a
phase-field model without explicitly considering a surface \cite{Yabunaka:2012}.

Although we focus on an outer fluid, 
generalization of the
models to include
%describe 
the inner concentration 
%or both the outer and inner concentrations 
is straightforward.
%our approach is also applicable to
%an inner bulk of a droplet with a source of chemicals.
We may also consider production rather than consumption of surfactants at an interface,
in which case spontaneous motion is realized for $\gamma_c < 0$.
In fact, spontaneous motion has been observed for complexes of surfactants
and ions that exhibit lower surface activity than the surfactants alone \cite{Thutupalli2011}.

\begin{acknowledgments}
The authors are grateful to T. Ohta for helpful discussions.
KHN acknowledges the support of a fellowship from the JSPS (No.23-1819).
NY acknowledges the support by a Grant-in-Aid for Young Scientists
(B) (No.23740317).
%Part of the numerical calculations in this work was carried out on
% Altix3700 BX2 at YITP in Kyoto University.
\end{acknowledgments}

\appendix
\section{derivation of Eqs.(\ref{Lambda0})-(\ref{Lambda2})}

In this appendix, we give a detailed derivation of the coefficients $\Lambda_{ab}$ in
the amplitude equations (11) and (12).
The dimensional analysis show that the coefficients have the dimension
of length; they are functions of $\lambda$ and $R$.
Introducing length and time scales, $L$ and $\tau$, the parameters are
scaled as
%\begin{align}
 $D  \sim L^2/\tau$, 
$\alpha \sim 1/L $, 
$A_1 \sim 1/L^2$, 
$u = u_1 A_1 \sim L/\tau$, and
$u_1 \sim L^3/\tau$
.
%\end{align}
Then the coefficients of amplitude equations are expressed as 
%\begin{align}
$\Lambda_{02} \sim L^0$, 
$\Lambda_{03} \sim L $,
$ \Lambda_{11} \sim 1/L$,
$\Lambda_{12} \sim L^0$, 
$\Lambda_{13} \sim L $, and
$\Lambda_{14} \sim L^2$
.
%\end{align}
Using the coefficients, the steady velocity of a droplet is obtained from (12) as
\begin{align}
 u = u_1 A_1 =
\sqrt{
\frac{- \left( \frac{D_s}{R^2} + \kappa_s
+ \frac{D \alpha}{\lambda} 
\right)
+ \Lambda_{12} u_1 (c_{\infty} - \alpha A_0)
}
{\Lambda_{13}\frac{\alpha}{D} 
+ \Lambda_{14} \frac{u_1}{D^2} (c_{\infty} - \alpha A_0)}
}.
\end{align}
Later, we will find $\Lambda_{12} \simeq \lambda/R$ and $\Lambda_{14}
\simeq \lambda^4/R^2$ which leads to the characteristic droplet velocity
under $\alpha \ll \kappa_s \lambda/D$  as
\begin{align}
 u \simeq u_0 = \sqrt{\frac{D \kappa R}{\lambda}}.
\end{align}
In order to obtain the concrete form of the coefficients, the Helmholtz
equation with nonlinear advection is solved neglecting time derivative
in (6),
\begin{align}
\nabla^2 c - \frac{1}{\lambda^2} (c-c_{\infty}) 
&=
\frac{{\bf v} \cdot \nabla c}{D},
\label{Helmholtz}
\end{align}
where the velocity field in the co-moving frame with the droplet is given %in the literature though it is given
explicitly here as \cite{levan:1981,Kitahata:2011}
\begin{align}
 v^o_r (r,\theta) 
& = 
- 
 u \left( 1 - \frac{R^3}{r^3} \right)
 P_1 (\cosT)
\nonumber \\
&- \sum_{n=2}^{\infty}
\frac{n(n+1)}{2n+1} u_n A_n
 \left[
\left( \frac{R}{r} \right)^n 
- \left( \frac{R}{r}\right)^{n+2} 
\right]
P_{n} (\cosT),
\label{velocity.r}
\end{align}
\begin{align}
 v^o_{\theta} (r,\theta) 
& =
- u \left( 1 + \frac{R^3}{2 r^3} \right)
 \odiff{P_1}{\theta} (\cosT)
\nonumber \\
&- \sum_{n=2}^{\infty}
\frac{u_n A_n}{2n+1}
 \left[
\frac{(n-2)R^n}{r^{n}} - \frac{n R^{n+2}}{r^{n+2}} 
\right]
\odiff{P_{n}(\cosT)}{\theta},
\end{align}
\begin{align}
 v^i_r (r,\theta) 
& =
- \frac{3}{2} u \left[
\left(\frac{r}{R}\right)^2 - 1
\right] \cosT
\nonumber \\
&- \sum_{n=2}^{\infty}
\frac{n(n+1)}{2n+1}
 u_{n} A_n \left[
\left( \frac{r}{R} \right)^{n+1}
- \left( \frac{r}{R} \right)^{n-1}
\right] P_n (\cosT),
\end{align}
\begin{align}
 & v^i_{\theta} (r,\theta) 
\nonumber \\
& =
- \frac{3}{2} u \left[
2 \left(\frac{r}{R}\right)^2 - 1
\right] \odiff{P_1}{\theta}
\nonumber \\
&-\sum_{n=2}^{\infty}
\frac{u_n A_n}{2n+1}
 \left[
(n+3) \left( \frac{r}{R} \right)^{n+1}
- (n+1) \left( \frac{r}{R} \right)^{n-1}
\right] \odiff{P_n (\cosT)}{\theta},
\end{align}
where $P_n (\cosT)$ is the $n$th-degree Legendre polynomial and
\begin{align}
 u_n = 
- \frac{\gamma_c}{2(\eta_i + \eta_o)}.
\end{align}
Near the critical point of drift bifurcation, the velocity of the
droplet is small and accordingly the advection term is small.
The solution is expanded perturbatively as $c = c_{\infty} + c^{(0)} +
c^{(1)} + c^{(2)} + \cdots$ and at each order
(\ref{Helmholtz}) becomes 
\begin{align}
 D_o \nabla^2 c^{(0)} - \kappa_o c^{(0)}
&= 0
\label{Helmholtz_zeroth}
\end{align},
\begin{align}
 D_o \nabla^2 c^{(1)} - \kappa_o c^{(1)}
&=
\frac{{\bf v}^o \cdot \nabla c^{(0)}}{D},
\label{Helmholtz_1st}
\end{align}
\begin{align}
 D_o \nabla^2 c^{(2)} - \kappa_o c^{(2)}
&=
\frac{{\bf v}^o \cdot \nabla c^{(1)}}{D}
\label{Helmholtz_2nd}
\end{align}
 for the order of $\epsilon^2$, and so on.
Note that although we focus only on the outer concentration field, the inner
concentration field yields essentially the same equations.
Hereafter, we drop the subscript ``o''.

%\section{uniform state}
The solution of the zeroth-order equation (\ref{Helmholtz_zeroth})
satisfying the boundary condition (3) is given in (9) using $n$th-order
modified spherical Bessel function of the second kind $k_n(x) =
\sqrt{2/(\pi x)}\mathcal{K}_{n+1/2}(x)$ where $\mathcal{K}_n(x)$ is the
$n$th-order modified Bessel function of second kind \cite{arfken:1968}.
%\section{weakly nonlinear analysis}
%In this section, the detail calculation of $\Lambda_{ab}$ will be shown
At the first order in the expansion, we will solve
\begin{align}
  \nabla^2 c^{(1)} - \frac{1}{\lambda^2} c^{(1)} = 
- \frac{u_{1} A_1}{D} \left( 1 - \frac{R^3}{r^3} \right)
\pdiff{c^{(0)}}{r} P_1 (\cosT).
\label{Expansion.first}
\end{align}
This equation is the form of the inhomogeneous Helmholtz equation:
\begin{align}
  \nabla^2 \psi - \frac{1}{\lambda^2} \psi = - f(r,\theta),
\end{align}
which corresponds to $\psi = c^{(l)}$ and $f = f^{(l)}$ using $l$-th order expansion in (\ref{Helmholtz_1st}) and (\ref{Helmholtz_2nd}).
The inhomogeneous term is expanded as
\begin{align}
 f^{(l)}(r,\theta) = \sum_{n=0}^{\infty}
f^{(l)}_n (r) P_n (\cosT).
\end{align}
The general solution yields\cite{acrivos:1962}
\begin{align}
 \psi(r,\theta) = 
\sum_{n=0}^{\infty} \psi_n k_n (r/\lambda)
P_n (\cosT)
+ \int G({\bf r,r'}) f(r,\theta) d^3 {\bf r},
\end{align}
where the Green's function satisfies\cite{arfken:1968}
\begin{align}
 \left( \nabla^2 - \frac{1}{\lambda^2} \right)
G({\bf r},{\bf r}') = - \delta ({\bf r} - {\bf r}').
\label{Green.definition}
\end{align}
For the Helmholtz equation in three dimensions, the Green's function is
given as
\begin{align}
 G({\bf r,r'}) = \frac{e^{-|{\bf r-r'}|/\lambda}}{4 \pi |{\bf r-r'}|}.
\label{Green.3D.Helmholtz}
\end{align}

The inhomogeneous term $f^{(1)}$ at the first order in expansion
(\ref{Expansion.first})  is expressed as
\begin{align}
 f^{(1)}_1 (r) =
  \frac{e^{-(r-R)/\lambda}}{r^2}
R \left( 1 - \frac{R^3}{r^3} \right) 
\frac{r+ \lambda}{\lambda}
A_1(c_{\infty} - \alpha A_0) \frac{u_{1}}{D},
\end{align}
and $f^{(1)}_n=0$ for $n \neq 1$.
The boundary condition (3) is given as
\begin{align}
 c^{(1)} (R,\theta) = \alpha A_1 P_1(\cosT).
\end{align}
The solution would be
\begin{align}
 c^{(1)} (r,\theta) 
= & 
\left[ \alpha A_1 -\mathcal{R}_1(R)
\right]
\frac{k_1(r/\lambda)}{k_1 (R/\lambda)}
P_1(\cosT)
\nonumber \\
&
+ \mathcal{R}_1(r) P_1(\cosT).
\end{align}
%where
%$\frac{k_1(r/\lambda)}{k_1 (R/\lambda)} = \frac{r+\lambda}{R + \lambda} 
%\left( \frac{R}{r} \right)^2
%e^{-(r-R)/\lambda} $.
%The concrete form of
where 
$\mathcal{R}_1(r)$ is
\begin{align}
 \mathcal{R}^{(1)}(r) 
=& 
 \frac{1}{\lambda} 
\left[
k_1 (r/\lambda) \int_R^{r} f_{1}(r') i_1 (r'/\lambda) r'^2 dr'
\right.
\nonumber \\
&\left.
+ i_1 (r/\lambda) \int_r^{\infty} f_{1}(r') k_1 (r'/\lambda) r'^2 dr'
\right]
.
\end{align}
 $i_n(x)$ is the $n$th-order modified spherical Bessel function of first kind
$ i_n(x) = \sqrt{\pi/(2x)} \mathcal{I}_{n+1/2}(x)$ using the
$n$th-order modified Bessel function of first kind $\mathcal{I}_{n}(x)$.
This function has a simple form
\begin{align}
 \mathcal{R}^{(1)}(R) =
 \frac{1}{\lambda} 
 i_1 (R/\lambda) \int_R^{\infty} f_{1}(r') k_1 (r'/\lambda) r'^2 dr'.
\label{R1}
\end{align}
The flux is expressed as
\begin{align}
  \pdiff{c^{(1)} (R)}{r} 
= &
\frac{\alpha  A_1}{\lambda} 
\frac{k'_1(r/\lambda)}{k_1 (R/\lambda)}
P_1(\cosT)
\nonumber \\
& + \left[ 
\frac{i'_1(R/\lambda)}{i_1 (R/\lambda)} 
-\frac{k'_1(R/\lambda)}{k_1 (R/\lambda)}  
\right]\frac{\mathcal{R}^{(1)} (R)}{\lambda}
P_1(\cosT).
\label{flux.first}
\end{align}
For $R \gg \lambda$ (\ref{R1}) becomes
%\footnote{
%Note that the lowest order term in $\lambda$ vanishes.
%Therefore, we expand the integral in
%(\ref{integral.modified.spherical.bessel.first.approx}) up to
%$\mathcal{O}(1/x^2)$.
%This arises from the fact that $f_1(R) = 0$, which is from $v_r(R) = 0$
%in addition to the fact that the expansion
%(\ref{integral.modified.spherical.bessel.first.approx}) does not depend
%on $l$.  
%}
\begin{align}
 \mathcal{R}^{(1)}(R) \simeq
 \frac{3}{8} \frac{\lambda^2}{R}
A_1 (c_{\infty} - \alpha A_0) \frac{u_{1}}{D},
\end{align}
and 
%using
%$
% \frac{i'_1(x)}{i_1 (x)} 
%\simeq
%1
%$
%and
%$
% \frac{k'_1(x)}{k_1 (x)} 
%\simeq
%- 1
%$,
we obtain the flux as
\begin{align}
 D \pdiff{c^{(1)} (R)}{r} 
\simeq
\left[
- \frac{\alpha D }{\lambda}
+  \frac{3}{4} \frac{\lambda}{R}
 (c_{\infty} - \alpha A_0) u_{1}
\right] A_1 P_1 (\cosT)
\label{flux.first}
.
\end{align}
The concrete form of $\mathcal{R}^{(1)}(r)$ is 
\begin{align}
 \mathcal{R}^{(1)} (r) 
& = 
\frac{u_1 A_1 e^{-(r-R)/\lambda}}{8D r^3}
\left[
2(r-R)^2 R (2r+R)
\right.
\nonumber \\
& \left.
+ 6 r (r-R) R \lambda
-3 r (r-2R) \lambda^2
- 3 r \lambda^3
\right](c_{\infty} - \alpha A_0),
\end{align}
and the concentration is
\begin{align}
 c^{(1)} (r,\theta) 
& =
\frac{A_1 R e^{-(r-R)/\lambda}}{4 r^3(R+\lambda)}
\left[
4 R \alpha r (r+\lambda)
\right.
\nonumber \\
& \left.
+ \frac{u_1}{D}
\left(
-3 r^2 R^2 - 3 r R^2 \lambda + 2 r^3 (R+\lambda)
+ R^3 (R+\lambda)
\right)
\right.
\nonumber \\
& \left.
\times
 (c_{\infty} - \alpha A_0)
\right]
P_1(\cosT),
\end{align}
which is shown in Fig.2(B).
Note that without the assumption of $R \gg \lambda$ the second term inside the bracket of (\ref{flux.first}) is
replaced by 
 $\frac{3 R \lambda}{4 (R+\lambda)^2} (c_{\infty} -\alpha A_0) u_1$,
 which is
always positive.
This implies that this
term destabilizes the stationary state irrespective of the value of
$\lambda$, that is, $\kappa$.

%\subsection{second order}
Calculation of the higher order terms is tedious but straightforward.
The second order term in bulk concentration field satisfies 
\begin{align}
 \left( \nabla^2 - \frac{1}{\lambda^2} \right) c^{(2)}(r,\theta)
= - f^{(2)} (r,\theta),
\end{align}
where
\begin{align}
 f^{(2)} (r,\theta) 
= &
 \frac{u_{1} A_1}{D} \left( 1 - \frac{R^3}{r^3} \right)
\pdiff{c^{(1)}}{r} P_1
\nonumber \\
& + \frac{u_{1} A_1}{D} \left( 1 + \frac{R^3}{2 r^3} \right)
\frac{1}{r} \pdiff{c^{(1)}}{\theta} \odiff{P_1}{\theta}.
\end{align}
This is decomposed as
\begin{align}
 f^{(2)} (r,\theta) = 
f^{(2)}_{0} (r) P_0 (\cosT)
+ f^{(2)}_{2} (r) P_2 (\cosT)
\end{align}
using
\begin{align}
 P_1 (\cosT) P_1 (\cosT) 
= \frac{1}{3} P_0 (\cosT) 
+ \frac{2}{3} P_2 (\cosT),
\end{align}
\begin{align}
 \odiff{P_1 (\cosT)}{\theta}  \odiff{P_1 (\cosT)}{\theta} =
\frac{2}{3} P_0 (\cosT) - \frac{2}{3} P_2 (\cosT).
\end{align}
Since we focus on the zeroth and first modes
and the boundary condition is
\begin{align}
 c^{(2)}(R)=0,
\end{align}
the general solution is expressed as
\begin{align}
 c^{(2)}(r,\theta) = &
- \mathcal{R}^{(2)} (R) 
\frac{k_0(r/\lambda)}{k_0(R/\lambda)} 
+ \mathcal{R}^{(2)}(r)
,
\end{align}
where
\begin{align}
 \mathcal{R}^{(2)} (r) 
& = 
\frac{1}{\lambda} 
\left[
k_0 (r/\lambda) \int_R^{r} f^{(2)}_{0}(r') i_0 (r'/\lambda) r'^2 dr'
\right.
\nonumber \\
& \left.
+ i_0 (r/\lambda) \int_r^{\infty} f^{(2)}_{0}(r') k_0 (r'/\lambda) r'^2 dr'
\right].
\end{align}
Similar to (\ref{flux.first}), the flux is expressed as
\begin{align}
 \pdiff{c^{(2)} (R)}{r} 
&=
\left[
\frac{i'_0(R/\lambda)}{i_0 (R/\lambda)}
-\frac{k'_0(R/\lambda)}{k_0 (R/\lambda)}
\right]\frac{\mathcal{R}^{(2)} (R)}{\lambda}
P_0(\cosT)
\nonumber \\
& \simeq
\frac{2 \mathcal{R}^{(2)} (R)}{\lambda}
P_0(\cosT)
.
\end{align}
with
\begin{align}
\mathcal{R}^{(2)} (R) 
& = 
\frac{u_1 A_1^2 \alpha \lambda^2}{8 D (R+\lambda)}
+ 
\frac{u_1^2 A_1^2}{480 D^2 \lambda^4 (R+\lambda)}
\nonumber \\
& \times
\left[
 \lambda \left(
4 R^6 + 2 R^5 \lambda
- R^3 \lambda^3
+ 3 R^2 \lambda^4
- 9 R\lambda^5
+ 30 \lambda^6
\right)
\right.
\nonumber \\
& \left.
- 8 e^{2R/\lambda}
R^6(R+\lambda)
\Gamma[2R/\lambda]
\right] (c_{\infty} - \alpha A_0).
\label{third.R.R}
\end{align}
$\Gamma[x]$ is the Gamma function.
Note that the concentration at this order is uniform since the coupling
of two $A_1$ modes results in $A_0$ mode.
%In this calculation of (\ref{third.R.R}), the exponential integral
%$Ei(2R/\lambda)$ appears.
For $R \gg \lambda$, it is known that expansion does not converge \cite{arfken:1968}.
Nevertheless, truncation at finite terms in the series of expansion
gives better approximation.

%\subsection{third order}
The similar calculation is applied for the third-order equation:
\begin{align}
 \left( \nabla^2 - \frac{1}{\lambda^2} \right) c^{(3)}(r,\theta)
= - f^{(3)} (r,\theta),
\end{align}
where
\begin{align}
 f^{(3)} (r,\theta) =
 \frac{u_{1} A_1}{D} \left( 1 - \frac{R^3}{r^3} \right)
\pdiff{c^{(2)}}{r} P_1 (\cosT)
.
\end{align}
The solution is expressed as
\begin{align}
 c^{(3)} (r,\theta) = &
\left[
- \mathcal{R}_3(R)
\frac{k_1 (r/\lambda) }{k_1(R/\lambda)}
+  \mathcal{R}_3(r)
\right]  P_1(\cosT)
,
\end{align}
where
\begin{align}
\mathcal{R}^{(3)}(r) 
= &
 \frac{1}{\lambda} 
\left[
k_1 (r/\lambda) \int_R^{r} f^{(3)}_{1}(r') i_1 (r'/\lambda) r'^2 dr'
\right.
\nonumber \\
& \left.
+ i_1(r/\lambda) \int_r^{\infty} f^{(3)}_{1}(r') k_1 (r'/\lambda) r'^2 dr'
\right],
\end{align}
with
\begin{align}
  f^{(3)} (r,\theta) &= 
f^{(3)}_{1} (r) P_1 (\cosT)
+ f^{(3)}_{3} (r) P_3 (\cosT).
\end{align}
The flux is calculated as
\begin{align}
 \pdiff{c^{(3)} (R)}{r} 
& =
\left[
\frac{i'_1(R/\lambda)}{i_1 (R/\lambda)}
-\frac{k'_1(R/\lambda)}{k_1 (R/\lambda)}
\right]\frac{\mathcal{R}^{(3)} (R)}{\lambda}
P_1(\cosT)
\nonumber \\
& \simeq
\frac{2 \mathcal{R}^{(3)} (R)}{\lambda}
P_1(\cosT)
\end{align}
with
\begin{align}
&\mathcal{R}^{(3)} (R) 
\nonumber \\
& \simeq 
\frac{3 u_1^2 A_1^3 \alpha \lambda^4 (R-\lambda) (R^2-3R\lambda + 3\lambda^2)
(2R^2 + 6 R\lambda + 3 \lambda^2)^2}{80 D^2 R^6 (R+\lambda)(R^2 + 3R
 \lambda + 3 \lambda^2)}
\nonumber \\
& - 
 \frac{u_1^3 A_1^3 R^{9}}{240  D^3 \lambda^5 (R+\lambda)} 
\left[ 
\mathcal{C}_1 e^{-2R/\lambda}
- \mathcal{C}_2 \Gamma [2R/\lambda]
\right]
(c_{\infty} -\alpha A_0),
\label{R3}
\end{align}
where
\begin{align}
\mathcal{C}_1 
\simeq &
\frac{877}{19305} 
+ \frac{6139}{38610} \frac{\lambda}{R}
+ \frac{26617}{4290} \left( \frac{\lambda}{R} \right)^2
+ \frac{541417}{15444} \left( \frac{\lambda}{R} \right)^3
\nonumber \\
&+ 94 \left( \frac{\lambda}{R} \right)^4
+ \cdots 
\end{align}
and
\begin{align}
\mathcal{C}_2
\simeq &
\frac{7016}{19305}
+ \frac{1754 }{19305 } \frac{R}{\lambda}
+ \frac{80728}{6435}  \frac{\lambda}{R}
+ \frac{490814}{6435} \left(\frac{\lambda}{R}\right)^2
\nonumber \\
& + 220  \left(\frac{\lambda}{R}\right)^3
+ 312 \left(\frac{\lambda}{R}\right)^4
+ \cdots.
\end{align}
We have used the integral including the Gamma function
\begin{align}
\int_r^{\infty}
r_1^n \Gamma [0,r_1/\lambda] d r_1 
&=
-\frac{r^{n+1}}{n+1}
\Gamma[0,r/\lambda]
+ \frac{\lambda^{n+1}}{n+1}
\Gamma[n+1,r/\lambda]
\end{align}
for $n \neq -1$.
In the limit of $\lambda \rightarrow 0$, the second term of (\ref{R3})
becomes $1071 \lambda^6 u_1^3 A_1^3/(128D^3 R)$.
For the finite
value of $\lambda$, as mentioned above,
the number of terms necessary for better approximation of the Gamma
function depends on the value of $\lambda$.
For $R \gg \lambda \gtrsim 0.01R$, we have confirmed numerically (\ref{R3}) is well approximated by
\begin{align}
\mathcal{R}^{(3)} (R) \simeq
\frac{3 u_1^2 A_1^3 \alpha \lambda^4 }{20 D^2 R^2}
%- \frac{3 u_1^2 A_1^3 \alpha \lambda^4 (R-\lambda)}{64 D^2 R^2 (R+\lambda)}
- \frac{5 u_1^3 A_1^3 \lambda^5}{6  D^3 R^2} 
(c_{\infty} -\alpha A_0).
\end{align}

The solution of $c(r)$ is plugged into $D \partial c/\partial r$ in (4)
and we obtain the set of amplitude equations (11) and (12).
The coefficients are given as
\begin{align}
\Lambda_{02} 
&=
\frac{\lambda}{4(R+\lambda)},
\label{Lambda02}
\end{align}
\begin{align}
\Lambda_{03} 
&\simeq
\frac{15 \lambda^3}{32 R^2},
\end{align}
\begin{align}
\Lambda_{12} 
&=
\frac{3 R \lambda }{4(R+\lambda)^2}
,
\label{Lambda12}
\end{align}
\begin{align}
\Lambda_{13} 
&\simeq
\frac{3 \lambda^3 }{10 R^2}
,
\end{align}
\begin{align}
\Lambda_{14} 
&\simeq
\frac{5\lambda^4}{3R^2}
.
\label{Lambda14}
\end{align}

%\bibliographystyle{unsrt}
%\bibliography{/home/yoshinaga/home/research/references/yoshinaga}

%Merlin.mbs v4.21 2009-07-09.
%

\end{document}